\documentclass[12pt,preprint,modern]{aastex}

\usepackage{amssymb}
\usepackage{times}
\usepackage{verbatim}
\usepackage{color}
\usepackage{xspace}

\usepackage{savesym}
\savesymbol{tablenum}
\usepackage{siunitx}
\restoresymbol{SIX}{tablenum}

\newcommand{\aciss}   {{ACIS-S}\xspace}
\newcommand{\chan}    {{\it Chandra}\xspace}
\newcommand{\ciao}    {{CIAO}\xspace}
\newcommand{\cmmthree}{\,\mathrm{cm\mthree}}

\newcommand{\cmmtwo}  {\,\mathrm{cm\mtwo}}
\newcommand{\cmptwo}  {\,\mathrm{cm\ptwo}}
\newcommand{\eflux}   {\,\mathrm{ergs\,cm\mtwo\,s\mone}}
\newcounter{ion}     \newcommand{\eli}[2]  {\setcounter{ion}{#2}#1{~\sc\roman{ion}}}
\newcommand{\hetg}    {{HETG}\xspace}
\newcommand{\hetgs}   {{HETGS}\xspace}
\newcommand{\kev}     {\,\mathrm{keV}}
\newcommand{\kms}     {\,\mathrm{km\,s\mone}}
\newcommand{\ks}      {\,\mathrm{ks}}
\newcommand{\mang}    {\,\mathrm{{\mbox{\AA}}}\xspace}
\newcommand{\mk}      {\,\mathrm{MK}}
\newcommand{\mone}  {^{-1}}
\newcommand{\mtwo}  {^{-2}}
\newcommand{\mthree}{^{-3}}
\newcommand{\nh}      {{N_\mathrm{H}}\xspace}
\newcommand{\pc}      {\,\mathrm{pc}}

\newcommand{\ptwo}  {^2}

\newcommand{\tmax}    {{T_\mathrm{max}}\xspace}
\newcommand{\tmin}    {{T_\mathrm{min}}\xspace}
\newcommand{\xmm}   {{\it XMM-Newton}\xspace}
\newcommand{\zpuppis} {{$\zeta\,$Puppis}\xspace}
\newcommand{\zpup}    {{$\zeta\,$Pup}\xspace}

\newcommand{\Change}[1]{{\bf#1}}
\renewcommand{\Change}[1]{#1}  

\shorttitle{The Hottest Plasma in the $\zeta\,$Puppis Wind}
\shortauthors{Huenemoerder et al.}

\begin{document}

\title{A Deep Exposure in High Resolution X-Rays Reveals the Hottest Plasma in
  the $\zeta\,$Puppis Wind}

\author[0000-0002-3860-6230]{David P.\ Huenemoerder}
\affiliation{Massachusetts Institute of Technology ,
  77 Massachusetts Ave.,
  Cambridge, MA 02139, USA
}

\author[0000-0002-7204-5502]{Richard Ignace}
\affiliation{Department of Physics \& Astronomy,
  East Tennessee State University,
  Johnson City, TN 37614 USA}

\author{Nathan A.\ Miller}
\affiliation{Department of Physics and Astronomy ,
  University of Wisconsin--Eau Claire,
  Eau Claire, WI 54701 USA
}

\author{Kenneth G.\ Gayley}
\affiliation{Department of Physics and Astronomy ,
  University of Iowa,
  Iowa City, IA 52242  USA
}

\author{Wolf-Rainer Hamann}
\affiliation{Institute for physics and astronomy, 
  University of Potsdam,
  Karl-Liebknecht-Str. 24/25, 14476,
  Potsdam, Germany
}

\author{Jennifer Lauer} 
\affiliation{Harvard-Smithsonian Center for Astrophysics,
  60 Garden Street,
  Cambridge, MA 02138, USA
}

\author[0000-0002-4333-9755]{Anthony~F.~J.~Moffat}
\affiliation{ D\'epartement de Physique,
  Universit\'e de Montr\'eal,
  C. P. 6128, succ. centre-ville,
  Montr\'eal (Qc) H3C 3J7, 
  and Centre de Recherche en Astrophysique du Qu\'ebec, Canada
}

\author[0000-0003-4071-9346]{Ya\"el Naz\'e 
}
\affiliation{F.R.S.-FNRS Senior Research Associate, Groupe d'Astrophysique des Hautes Energies, STAR,
  Université de Liège,
  Quartier Agora (B5c, Institut d'Astrophysique et de Géophysique),
  Allée du 6 Août 19c, 4000 Sart Tilman,
  Liège, Belgium     
}

\author[0000-0003-3298-7455]{Joy S.\ Nichols}
\affiliation{Harvard-Smithsonian Center for Astrophysics,
  60 Garden Street,
  Cambridge, MA 02138, USA
}

\author[0000-0003-0708-4414]{Lidia Oskinova}
\affiliation{Institute for physics and astronomy, 
  University of Potsdam,
  Karl-Liebknecht-Str. 24/25, 14476,
  Potsdam, Germany
}

\author[0000-0002-2806-9339]{Noel D.\ Richardson}
\affiliation{Embry-Riddle Aeronautical University,
  3700 Willow Creek Rd ,
  Prescott, AZ 86301, USA
}

\author{Wayne Waldron} 
\affiliation{Eureka Scientific, Inc.,
  2452 Delmer Street,
  Oakland, CA 94602, USA
}
\begin{abstract}

We have obtained a very deep exposure ($813\,\mathrm{ks}$) of $\zeta\,$Puppis
(O4 supergiant) with the {\it Chandra} HETG Spectrometer.  Here we report on
analysis of the $1$--$9\,\mathrm{\AA}$ region, especially well suited
for {\it Chandra}, which has a significant contribution from continuum
emission between well separated emission lines from high-ionization
species.  These data allow us to study the hottest plasma present
through the continuum shape and emission line strengths.  Assuming a
powerlaw emission measure distribution which has a high-temperature
cut-off, we find that the emission is consistent with a thermal
spectrum having a maximum temperature of $12\,\mathrm{MK}$ as
determined from the corresponding spectral cut-off.  This implies an
effective wind shock velocity of $900\,\mathrm{km\,s^{-1}}$, well
below the wind terminal speed of $2250\,\mathrm{km\,s^{-1}}$.  For
X-ray emission which forms close to the star, the speed and X-ray flux
are larger than can be easily reconciled with strictly self-excited
line-deshadowing-instability models, suggesting a need for a fraction
of the wind to be accelerated extremely rapidly right from the base.
This is not so much a dynamical instability as a nonlinear response to
changing boundary conditions.

\end{abstract}




\section{Introduction}\label{sec:intro}

\zpuppis is one of the optically brightest O-stars, and is close  at an
estimated distance of only $332\pc$ \citep{Howarth:vanLeeuwen:2019}.
As such, this O4If(n)p star \citep{Sota:Apellaniz:Walborn:al:2011} has
been the target of many important studies to characterize its
variability and properties.  Much of this work was summarized in a
recent analysis of optical spectroscopy and high-precision photometry
by \citet{ramiar:al:2018}, to whom we refer readers for relevant
background discussion about the target star.

\zpup's relatively high X-ray flux of about $10^{-11}\eflux$
($1$--$40\mang$) allows detailed analysis of its spectrum when
dispersed by the High Energy Transmission Grating Spectrometer (HETGS)
aboard the \chan X-ray Observatory.  An existing $68\ks$ \chan/\hetgs
spectrum of \zpup was analyzed by \citet{Cassinelli:Miller:al:2001}
who demonstrated that the spectrum is well described by
multi-temperature thermal plasma models.  To take advantage of the
HETGS high resolution by driving down the statistical errors, we have
obtained an additional 813 kiloseconds of exposure time for this star.

Below about $9\mang$, the resolution of the \hetgs and the low
emission line density allows a relatively clean separation between
lines and continuum.  The thermal Bremsstrahlung (free-free) continuum
shape is sensitive to the hottest plasmas expected ($\sim10\mk$).  In
addition, the emission lines in this region are from highly ionized
species (H- and He-like ions) of relatively abundant elements (Mg, Si,
S, Ar, Ca, Fe), and are also formed at high temperatures.

The \xmm spacecraft has been used to study this star, starting with
\citet{kahn2001} and continuing with the extensive series of papers
\citep[e.g.][]{naze:flores:rauw:2012, naze:al:2018}.
However, the sensitivity of the RGS instrument aboard the \xmm
spacecraft is much lower below $9\mang$ and its lower
resolution impedes the clear separation of the continuum and emission
lines.  Hence, \hetg is the best current instrument for study of both
the line and continuum emission in the short wavelength region we
concentrate on here.

In this paper, our primary interest is in determinateion of plasma
temperatures, in particular the characterization and measurement of
high-temperature lines and continuum in the \chan/\hetgs spectrum of
\zpup.  If X-ray emission originates solely from embedded-wind shocks
\citep{Lucy:White:1980, feldmeier:puls:al:1995}, then the maximum
temperature is determined by the highest relative velocities of the
colliding structures, and the amount of emission is indicative of the
emitting volume.  Hot thermal plasmas emit strongly in bound-bound
lines of highly ionized states, \Change{in thermal Bremsstrahlung
  radiation, and in bound-free continuum emission.  Bremsstrahlung and
  bound-free emission drop exponentially for photon energies larger
  than about $kT$ of the plasma (in which $k$ is Boltzmann's constant
  and $T$ is the plasma temperature).  For some plasma temperatures,
  bound-free emission can exceed that from thermal Bremsstrahlung at
  high energies \citep{landi:2007,kaastra:paerels:al:2008}. However,
  the continuum shape is similar to that from Bremsstrahlung.}  The
the drop in the spectrum at high energies (short wavelengths) is
sensitive to the hottest temperatures present in a multi-thermal
plasma.  Thermal emission lines are also very sensitive to
temperature, having temperatures of peak emissivities that increase
with atomic number; they typically emit most strongly over a range in
temperature of about 0.3 dex (defined by the full-width-half-maximum
of their emissivity curves).  Hence, we concentrate here on the
2--9$\mang$ region due to the relative sparseness of emission lines,
the high temperature sensitivity, and the reduced importance of wind
absorption.  \Change{The temperature sensitivity occurs because for
  the expected plasma temperatures from embedded wind shocks
  ($2$--$10\mk$), the spectrum will drop accordingly somewhere in the
  $1$--$10\mang$ range.  In this regime, the wind absorption is
  relatively unimportant since the continuum opacity (due mainly to
  K-shell absorption by metals) drops very steeply to short
  wavelengths, so we do not have to be concerned with the wind density
  structure to interpret the spectral shape.}
  
This new dataset allows us to probe the highest temperature plasma
with unprecedented precision, providing a basic test for any models of
X-ray production.  Our aim is to use simple but robust methods to
infer this maximum temperature.

\section{Observations, Calibration, and Analysis}\label{sec:obscal}

The observations analyzed here are part of a Cycle 19 Large Project
(PI Waldron, Sequence Number 201176), to which we also added the first
\hetgs spectrum from 2000 (PI Cassinelli, Observation ID 640).  \chan
Observations were made with the \hetg in conjunction with the \aciss
detector array, a configuration known as \hetgs, the High Energy
Transmission Grating Spectrometer \citep{HETG:2005}.  An observing log
is given in Table~\ref{tbl:obs}.  We note that secular changes in the
\hetgs sensitivity have largely occurred at longer wavelengths than
those we are concerned with here \Change{(less than 10\% change below
$5\mang$)}, making it relatively easy to seamlessly include the ObsID
640 data with the more recent data.\footnote{See the \chan
  data analysis web-page entry on {\it ACIS QE Contamination}
  (\url{https://cxc.harvard.edu/ciao/why/acisqecontamN0011.html}), and
  the {\it \chan Proposers' Observatory Guide} \S 6.5.1: {\it Molecular
  Contamination of the OBFs}
  (\url{https://cxc.harvard.edu/proposer/POG/html/chap6.html\#tth\_sEc6.5.1}).
}  
Any remaining small changes in
the instrument's properties are dealt with through calibration, as
described below.

Spectra were extracted using standard \ciao procedures \citep[version
  4.11;][]{CIAO:2006}, using calibration database version 4.8.2.  For
each of the 22 observations, we extracted the positive and negative
first diffraction-order count-spectra for the two types of gratings
(High and Medium Energy Gratings, or HEG and MEG).  For each of these
count spectra, responses were computed from \ciao programs---the
effective area files, or ``ARFs'', and the response matrices, or
``RMFs''.  The latter encodes the energy-dependent line-spread
function and enclosed count fraction for the extraction aperture
used.

The variability in the $2$--$9\mang$ spectral region is small enough
to justify our analysis of the overall mean spectrum (see
Section~\ref{sec:xrlc}).  Detailed analysis of the X-ray variability
of this star will be described in a separate paper (Nichols, et al.,
in preparation).

While background is usually negligible for \hetgs\footnote{For
  information on the HETGS background rate, see the relevant section
  of the {\it \chan Proposers' Observatory Guide},
  \url{https://cxc.harvard.edu/proposer/POG/html/chap8.html\#tth_sEc8.2.3}.}
due to the ability to do spectral order-sorting through rejection of
events using the intrinsic energy resolution of the detector,
background does become appreciable in our analysis in the
1.7--3.0$\mang$ region where the source flux and the effective area
both drop (the effective area at $3\mang$ is about $75\cmptwo$ but is
only $18\cmptwo$ at $1.8\mang$).  The interesting high-temperature
lines of \eli{Fe}{25}, $\lambda 1.85\mang$ (maximum emissivity at
$\sim60\mk$) occur in this region, along with the continuum from the
highest temperature plasma.  Hence, we have included background
spectra, taken from regions adjacent to the dispersed spectra, in our
analysis.  This is always done by adding background-count to
model-count spectra during fitting; the only time we subtract
background counts from the data is for the purposes of visualization.

Fitting and modeling were done using the {\it Interactive Spectral
  Interpretation System}
\citep[ISIS\footnote{\url{http://space.mit.edu/cxc/software/isis}},][]{Houck:00}
which provides interfaces to an atomic database
\citep[``AtomDB'';][]{Foster:Smith:Brickhouse:2012} for atomic data
and emissivities used to construct plasma models \Change{(including bound-free,
free-free, two-photon, and dielectronic recombination processes); it
provides access to Xspec models, from which we used
``{\tt phabs}'' for absorption, which is based on cross-sections
  from \citet{verner:al:1996}.  This absorptions model is adequate for
  the spectral region of interest since the major sources of opacity
  are inner-shell electrons of high $Z$ ions.  The ionization state
  and non-Solar abundances of C, N and O do not matter below
  $\sim10\mang$.}
ISIS also has features
which facilitate simultaneous modeling of the 88 spectra (22
observations with 4 orders each) and their associated responses.

The fundamental paradigm in X-ray spectral analysis is iterative
forward folding, in which the instrumental response is applied to a
model spectrum via an integral equation in order to generate model
counts.  A background (empirical or model) is optionally added to the
model counts.  The model parameters are adjusted (for fitting done
here, usually with an amoeba-subplex optimization method) to minimize
the residuals between the observed and model counts, using Poisson
counting statistics.  This is necessary because while the response
matrix is nearly diagonal for diffraction grating instruments, it
still has significant off-diagonal terms (representing the spectral
resolution and instrumental profile), and cannot be inverted uniquely.
The only robust method is to maintain each count-spectrum and their
individual reponses.\footnote{If unfamiliar with forward-folding of a
  model through the response, see the discussion in Section~7.7 of the
  ISIS manual at
  \url{https://space.mit.edu/cxc/software/isis/manual.pdf}, which is
  germane to both low and high resolution spectroscopy, with single or
  multiple diffraction orders.}  We do use an approximate
``unfolding'' for visualization in division of the counts by the model
count rate per unit flux.  This does not deconvolve line profiles, nor
is it guaranteed to be accurate in regions of rapidly changing
spectral response with wavelength.  But it is good for visualization
of the data in model space and for assessing the overall quality of
the fit (residuals are still formed between modeled and observed
counts).  To improve efficiency in fitting, we combine datasets across
observations and orders for each grating type; that is, we combine HEG
positive and negative orders, and likewise for MEG.  This means that
models only need to be evaluated twice, once for each grid.
Internally to the system, however, each dataset and response are still
unique: model counts are evaluated for each array, then counts are
summed, models are summed, and residuals computed.  Hence, we can work
with any subset of the 22 observations without creating actual files
(counts and responses) for each permutation of interest.

%
\begin{deluxetable}{rrc}
  \tablecaption{Observations Analyzed}
  \tablehead{
    \colhead{ObsID}&
    \colhead{Observation Start Time}&
    \colhead{Exposure}\\
    &&\colhead{[ks]}
  }
  \startdata
  640& 2000-03-28T13:31:41& 67.7\\
  21113& 2018-07-01T20:18:49& 17.7\\
  21112& 2018-07-02T22:57:54& 29.7\\
  20156& 2018-07-03T16:06:38& 15.5\\
  21114& 2018-07-05T17:00:36& 19.7\\
  21111& 2018-07-06T05:00:09& 26.9\\
  21115& 2018-07-07T03:17:11& 18.1\\
  21116& 2018-07-08T02:20:58& 43.4\\
  20158& 2018-07-30T22:36:40& 18.4\\
  21661& 2018-08-03T11:42:46& 96.9\\
  20157& 2018-08-08T23:32:35& 76.4\\
  21659& 2018-08-22T02:13:29& 86.3\\
  21673& 2018-08-24T18:52:10& 15.0\\
  20154& 2019-01-25T03:21:34& 47.0\\
  22049& 2019-02-01T00:55:26& 27.7\\
  20155& 2019-07-15T00:04:38& 19.7\\
  22278& 2019-07-16T16:20:37& 30.5\\
  22279& 2019-07-17T14:52:40& 26.0\\
  22280& 2019-07-20T06:45:30& 25.5\\
  22281& 2019-07-21T21:13:28& 41.7\\
  22076& 2019-08-01T00:47:34& 75.1\\
  21898& 2019-08-17T03:16:06& 55.7\\
  \enddata
\end{deluxetable}\label{tbl:obs}
%

\section{Spectral Modeling}\label{sec:modeling}

We proceeded in iterative steps in modeling the spectral region below
$9\mang$.  Since emission lines are present which are formed at very
different temperatures (e.g, \eli{Mg}{11} and \eli{S}{16}), the plasma
is not iso-thermal, so we began by fitting a two-temperature-component
model to provide overall plasma characteristics and to guide further
analysis.  For our more definitive multi-temperature thermal plasmas,
each temperature component is constrained to have the same line shape,
elemental abundances, and multiplicative absorption function.
Because for the purposes of this paper we are primarily interested
only in the line strengths (as a probe of temperature, abundance, and
absorption), a Doppler shifted and broadened Gaussian is used to model
the emission line shapes.  A more detailed analysis of the lines would
include distributed wind absorption and emission in an expanding
medium \citep[see, e.g.][]{Owocki:Cohen:2001, Oskinova:al:2006,
  Zhekov:Palla:2007}, but as can be seen from the fits displayed
later, this simpler model is adequate for our purposes of modeling the
overall line fluxes and separating the lines from the continuum.
Whatever the actual physical effects determining the line shapes, it
is fortunate that the lines seem simply shifted rather than strongly
asymmetric \citep[see the fits in][]{Cassinelli:Miller:al:2001,
  Waldron:Cassinelli:2001}.  Likewise, the absorption is not really a
foreground screen of absorption; our assumption is similar to the
``exospheric'' approximation \citep{owocki:cohen:1999}, which assumes
all emission is from above the wind optical depth for X-ray emission,
$\tau_\lambda=1$.  This is reasonable for the shorter wavelengths
which can be seen from geometrically deep in the wind, as far down as
the photosphere \citep[see Figure 5 in][]{Oskinova:al:2006}. 
We add a ``foreground'' screen of absorption to approximate the
distributed wind absorption.  This should be reasonable in
understanding the shape of the spectral energy distribution.  In the
more detailed analysis of broad-band X-ray absorption of
\citet{leutenegger:cohen:al:2010}, it was found that below $10\mang$,
the foreground absorbing screen approximation was adequate.

The free parameters per plasma temperature component are then the
temperature and normalization (emission measure).  The line width and
shift were determined independently by fitting line-dense regions
\Change{($\sim9$--$13\mang$)} with an AtomDB model template, and then
frozen.  We used a uniform Doppler shift of $-450\kms$ and a Gaussian
broadening term (in addition to thermal broadening) with a full-width
half-maximum of about $1900\kms$, which is reasonable given the
terminal wind velocity of $2250\kms$ \citep{puls:markova:al:2006}.
The apparent Doppler shift is a signature of wind absorption, due to
line-of-sight absorption of the receding wind's X-ray emission.
\Change{In general, these parameters change with wavelength, due to the
  region of formation in the wind.  Using the mid-spectrum gives us an
  average value, and later during fitting we  bin the spectrum
  substantially so that we are not so sensitive to the specific
  values; we are primarily interested in the continuum shape and the
  integrated line fluxes as temperature diagnostics.}
We defer detailed line profile analysis to a follow-up paper (in
preparation).

Abundances of Mg, Si, S, and Ar were left free; for $2T$ fits, these
parameters are degenerate with $T$.  The result is a provisional model
which gives a reasonable match to the observed spectrum.  It required
temperatures of about $6.0$ and $14\mk$, with emission measures of
$2.6\times10^{55}$ and $2.3\times10^{54}\cmmthree$.

The absorption term required a column of $N_\mathrm{H} \sim
4\times10^{21}\cmmtwo$.  For comparison, the interstellar column is
only $2\times10^{20}\cmmtwo$, using
\Change{the color excess in the blue-to-visual color magnitudes,} $E(B-V)=0.04$
\citep{Howarth:vanLeeuwen:2019} and the standard dust-to-gas ratio
scaling relation, $N_\mathrm{H} \approx 5\times10^{21} \times
E(B-V)\cmmtwo$.  While this spectral region is less sensitive to wind
absorption than the overall X-ray spectrum, it is not negligible.  Our
model is not sufficient to apply to the entire HETG region, but since
we are mainly interested in the highest temperature, this is well
constrained (to about $0.1\,$dex).  Improvements to a global model
will come when line profiles are modeled in detail.

Using this provisional model, we evaluated the presence of
``line-free'' regions by taking the ratio of the total to the
continuum-only model count spectrum, and defined line-free as regions
where the deviation from the model continuum is below 5\%.  Over the
$2$--$9\mang$ range, about 45\% of the resolution elements are
line-free (at HEG or MEG resolution; the lines are resolved in each
channel).  Above $9\mang$, there are no such regions. Over
$2$--$9\mang$, 20\% of the photon flux is from the line-free regions,
whereas the complement 80\% of the region's flux is about equally
divided between line and continuum photons.

We attempted fits to just the line-free region, but the lower counts
made for much more uncertain parameters.  The hotter plasma continuum
only becomes appreciable below about 3--4$\mang$, where signal drops.
The lines of Mg, Si, S, Ar, Ca, and Fe are needed to constrain plasmas
having temperatures from 6--60$\mk$.  The lines contribute
significantly to the counts improving the overall statistics, and they
provide substantial leverage on temperature determination.

For a more physically motivated model than the 2-temperature fit, such as
might be encountered with an ensemble of shocks in a clumpy wind
\citep[see, for
  example][]{cassinelli:ignace:al:2008,ignace:waldron:al:2012}, we
have adopted a power-law differential emission measure ($DEM$) model
of the form,
\begin{equation}\label{eq:dem}
  dEM / dT  = [n_e n_H \, dV/dT] = D_0(r)\,T^{-\beta},
\end{equation}
where $dEM/dT$ is the differential emission measure,
\Change{$V$ is the volume of X-ray-emitting plasma,}
$D_0$ is a
normalization factor, $n_e$ and $n_H$ are the electron and hydrogen
number densities, respectively, $T$ is the plasma temperature which we
limit to a range $T_\mathrm{min}$, $T_\mathrm{max}$, and the exponent
$\beta$.  In Appendix~\ref{sec:appa} we derive an analytic solution
for the continuum spectral energy distribution in the limit of
$T_\mathrm{min} = 0$ \citep[also see][]{brown:1974}.

We have implemented a power-law $DEM$ in ISIS using a weighted sum of
AtomDB model spectra on an evenly spaced logarithmic temperature grid
with parameters $T_\mathrm{min}$, $T_\mathrm{max}$, the $DEM$ exponent
$\beta$, an overall normalization term (proportional to the volume
emission measure\Change{, but not identical to $D_0$ in
  Equation~\ref{eq:dem} because the stellar distance is also
  included}), the plasma model parameters shared by all
components (Gaussian broadening, Doppler shift, and abundances), with
multiplicative absorption.

Our free parameters are the normalization, $\beta$, $\tmin$, $\tmax$,
$\nh$, and abundances of the most significant species in the region,
Mg, Si, S, and Ar.  We have obtained fits with these free parameters,
but the statistic's minimum is somewhat complicated with degenerate or
poorly constrained parameters---better minima are often found when
additional searches are performed, such as for confidence limits on
parameters.  Hence, we 
use a Markov-Chain, Monte-Carlo method
\citep[an ISIS implementation of the ][Python code]{emcee:2013} to
search parameter space and form confidence contours.\footnote{The
  ISIS version is described at
  \url{https://www.sternwarte.uni-erlangen.de/wiki/index.php/Emcee}
  and is available as part of the {\tt isisscripts} package
  available at \url{https://www.sternwarte.uni-erlangen.de/isis/}. }
From these runs, we could easily see a strong degeneracy between
the normalization and the minimum temperature.  Since the minimum
temperature is not an interesting parameter (and is set to zero in
the theoretical derivation in Appendix~\ref{sec:appa}), we froze
$\tmin$ to the best fit value of $6.0\,$dex K; it was always nearly
a factor of 10 or more below $\tmax$.  Furthermore, the shape of
the spectrum, which is the primary diagnostic of the highest
temperature plasma, is not very sensitive to $\tmin$.  The fits
then produced closed confidence contours for the normalization,
and did not change other confidence limits.  This does introduce
an overall systematic uncertainty in the model, but not one which
affects our primary results.

\begin{figure}[!htb]
  \centering\leavevmode
  \includegraphics[width=0.98\columnwidth]{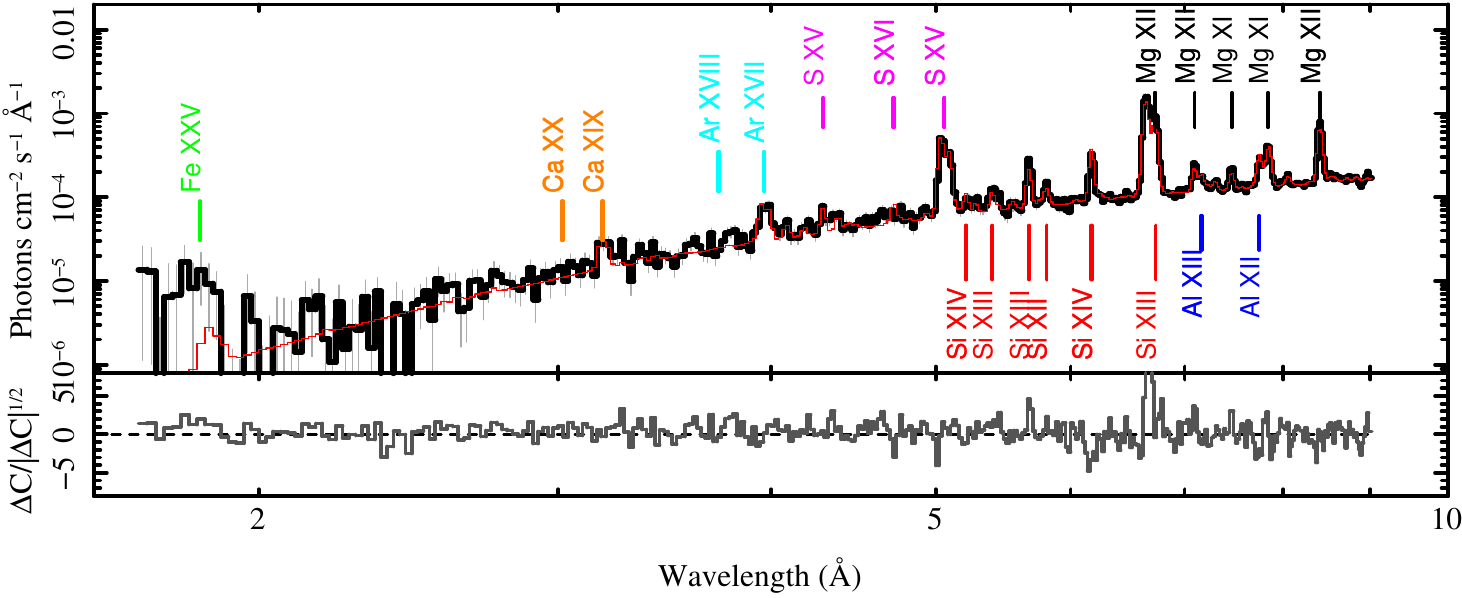}  
  \caption{X-ray spectrum (black) and absorbed powerlaw emission
    measure distribution' spectral model (thin red line).  The figure shows the combined
    HEG and MEG data for the entire $881\ks$ exposure.  Background has
    been subtracted from the data and model counts for plotting
    purposes.  Counts have been converted to approximate flux through
    division by the count rate per unit flux.  The lower panel shows
    the statistical residuals of data counts to model counts.  The
    large residuals in the region fit are from \eli{Si}{13}
    ($6.7\mang$) because we have not compensated for photoexcitation
    which suppresses the forbidden lines and enhances the
    intercombination lines (the forbidden-to-intercombination line
    ratios for helium-like ions will continue to be very interesting
    for this star, but their precise strengths are not important for
    the global fitting being performed here).  Positions of some lines
    have been marked (whether detected or not), using different colors
    for each element.}
  \label{fig:specoverview}
\end{figure}

Figure~\ref{fig:specoverview} shows the data, model, and residuals.
We only fit the spectrum below $9\mang$, and in a few near-continuum
bins near $13\mang$ which are important for constraining $N_\mathrm{H}$.  We
also ignored the \eli{Si}{13} intercombination and forbidden lines
near $6.7\mang$, since these are strongly affected by the
photospheric UV radiation field, as can be seen in the residuals
when we evaluate the model over this region.  (In our model, we also
set the Fe relative abundance to $0.5$, since this gave a better
extrapolation of our fit to the $10$--$12\mang$ region, but it is
otherwise not important to the spectral region fit.)

\begin{figure}[!htb]
  \centering\leavevmode
  \includegraphics[width=0.80\columnwidth]{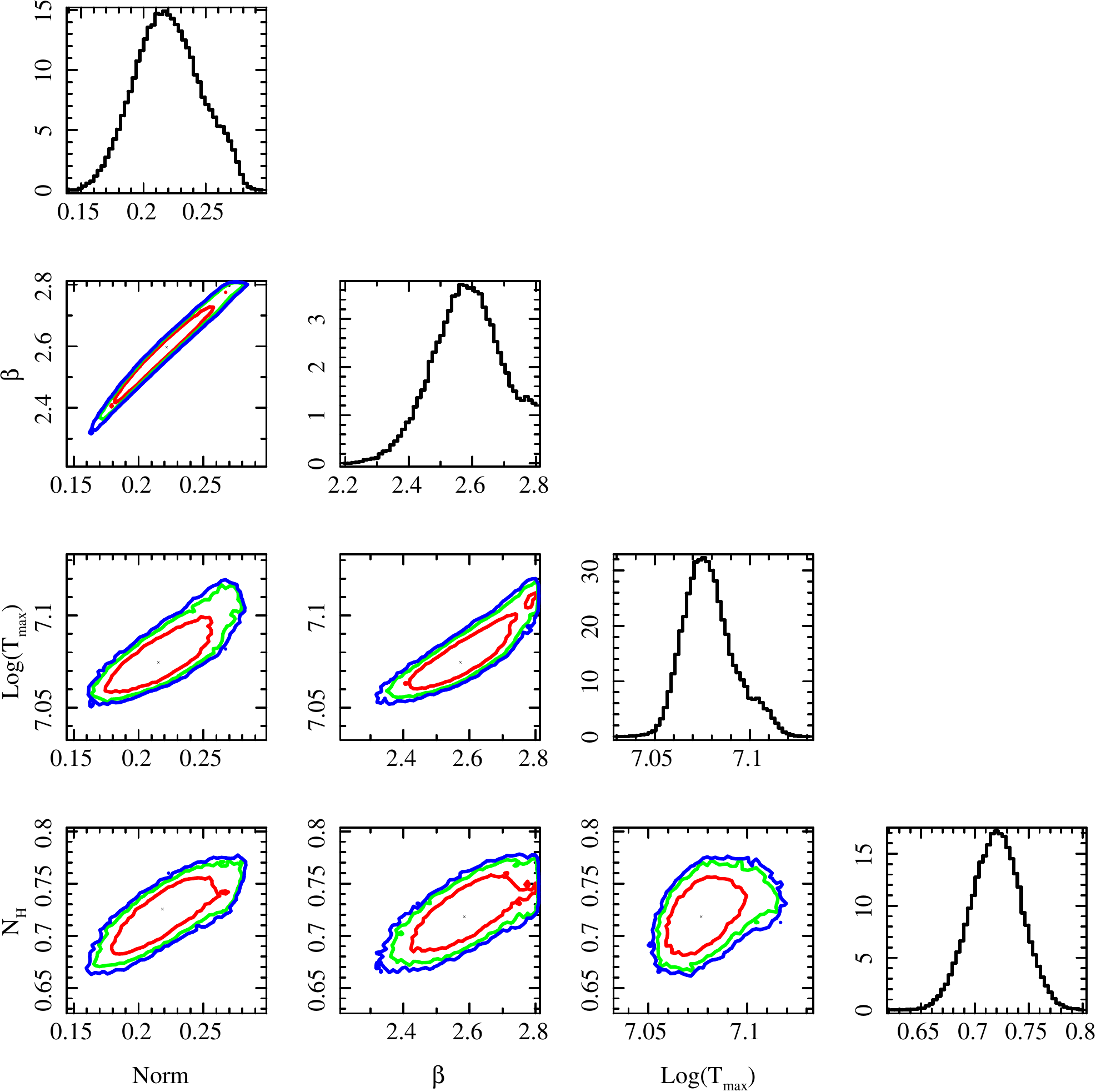}  
  \caption{Parameter-pair confidence contours from 68\% (inner, red),
    90\% (middle, green), and 95\% (blue, outer).  The $Norm$ is in
    units of $10^{-14} / (4\pi d^2) \times VEM$ (the volume emission
    measure), $\beta$ is the $DEM$ exponent, temperatures are in
    ($\log$) K, and $N_\mathrm{H}$ is in units of
    $10^{22}\cmmtwo$. The marginal histograms for each parameter are
    shown on the diagonal.}
  \label{fig:emceemain}
\end{figure}

\begin{figure}[!htb]
  \centering\leavevmode
  \includegraphics[width=0.98\columnwidth]{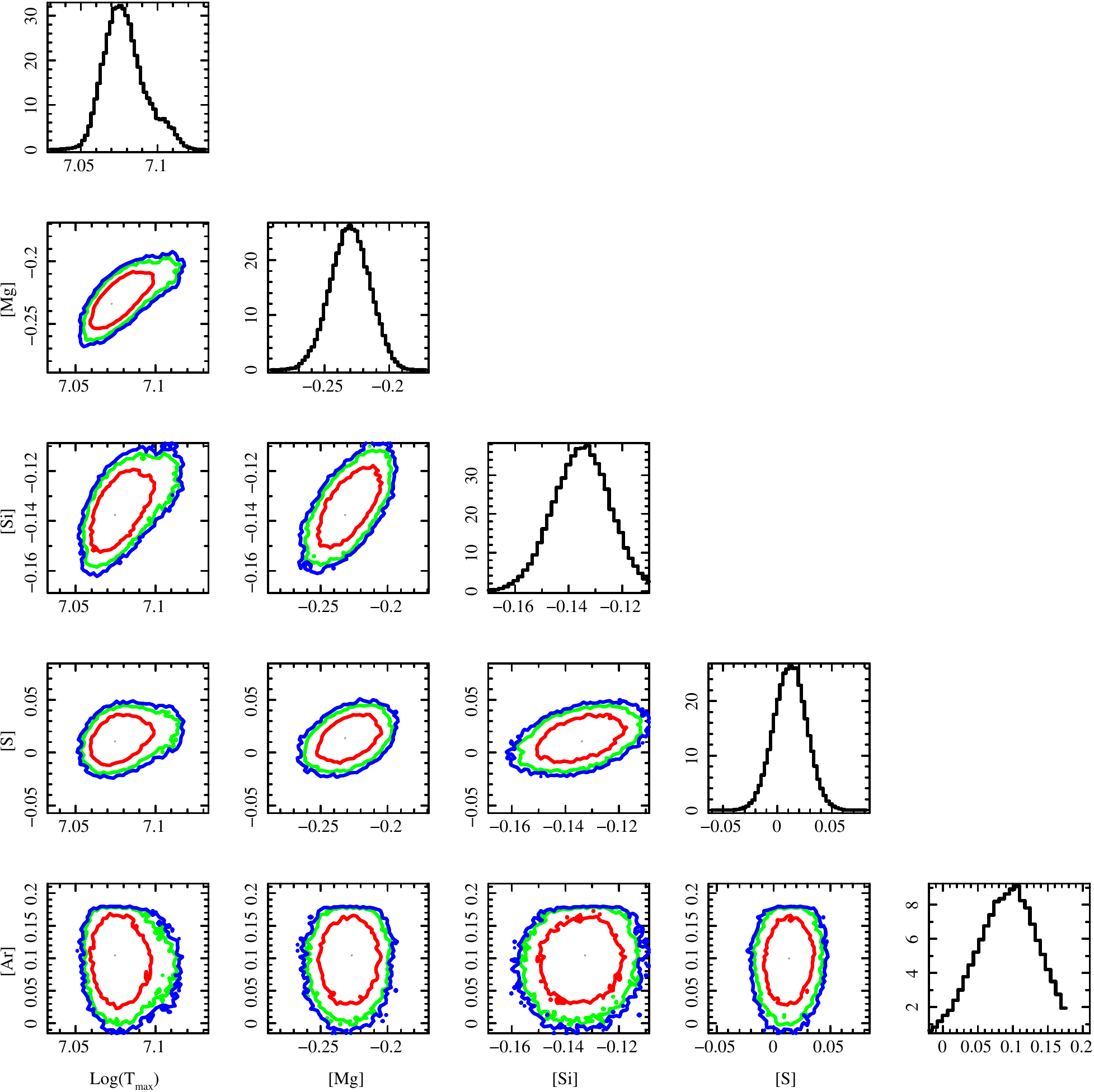}  
  \caption{The abundance parameter-pair confidence contours.
    Abundances are given in the logarithmic difference from the values
    in Table~\ref{tbl:abunds}. The marginal histograms for each
    parameter are shown on the diagonal. }
  \label{fig:emceeabunds}
\end{figure}

Figure~\ref{fig:emceemain} shows parameter-pair confidence contours
for the normalization, maximum temperature, absorbing column, and
differential emission measure exponent.  The relative abundance
parameter pair confidence contours are shown in
Figure~\ref{fig:emceeabunds}.  The reference abundances relevant to
the region fit are listed in Table~\ref{tbl:abunds}.

%
\begin{deluxetable}{rrrc}
  \tablecaption{Reference Abundances}
  \tablehead{
    \colhead{$Z$}&
    \colhead{Elem.}&
    \colhead{$\log N$}&
    \colhead{Mass Frac.}\\
    &
    & (dex)
    &
  }
  \startdata
  $ 1$&   H&   $12.00$&  \num{5.98E-01}\\
  $ 2$&  He&   $11.21$& \num{3.85E-01}\\
  $ 6$&   C&   $ 7.60$& \num{2.85E-04}\\
  $ 7$&   N&   $ 9.10$& \num{1.05E-02}\\
  $ 8$&   O&   $ 8.14$& \num{1.30E-03}\\
  $10$&  Ne&   $ 8.02$& \num{1.25E-03}\\
  $12$&  Mg&   $ 7.68$& \num{7.06E-04}\\
  $13$&  Al&   $ 6.54$& \num{5.55E-05}\\
  $14$&  Si&   $ 7.60$& \num{6.63E-04}\\
  $16$&   S&   $ 7.21$& \num{3.08E-04}\\
  $18$&  Ar&   $ 6.49$& \num{7.32E-05}\\
  $20$&  Ca&   $ 6.43$& \num{6.40E-05}\\
  $26$&  Fe&   $ 7.59$& \num{1.29E-03}\\
  \enddata
  \tablecomments{Abundances for H, He, C, N, and O are from
    \citet{bouret:hillier:al:2012}, and the remainder are from the
    Solar mass-fraction values of \citet{Asplund:Grevesse:al:2009}.}
\end{deluxetable}\label{tbl:abunds}
%

\section{X-Ray Variability of the Overall Flux Below 9 \AA}\label{sec:xrlc}

\begin{figure}[!htb]
  \centering\leavevmode
  \includegraphics*[width=0.30\columnwidth, viewport=0 0 430 650]{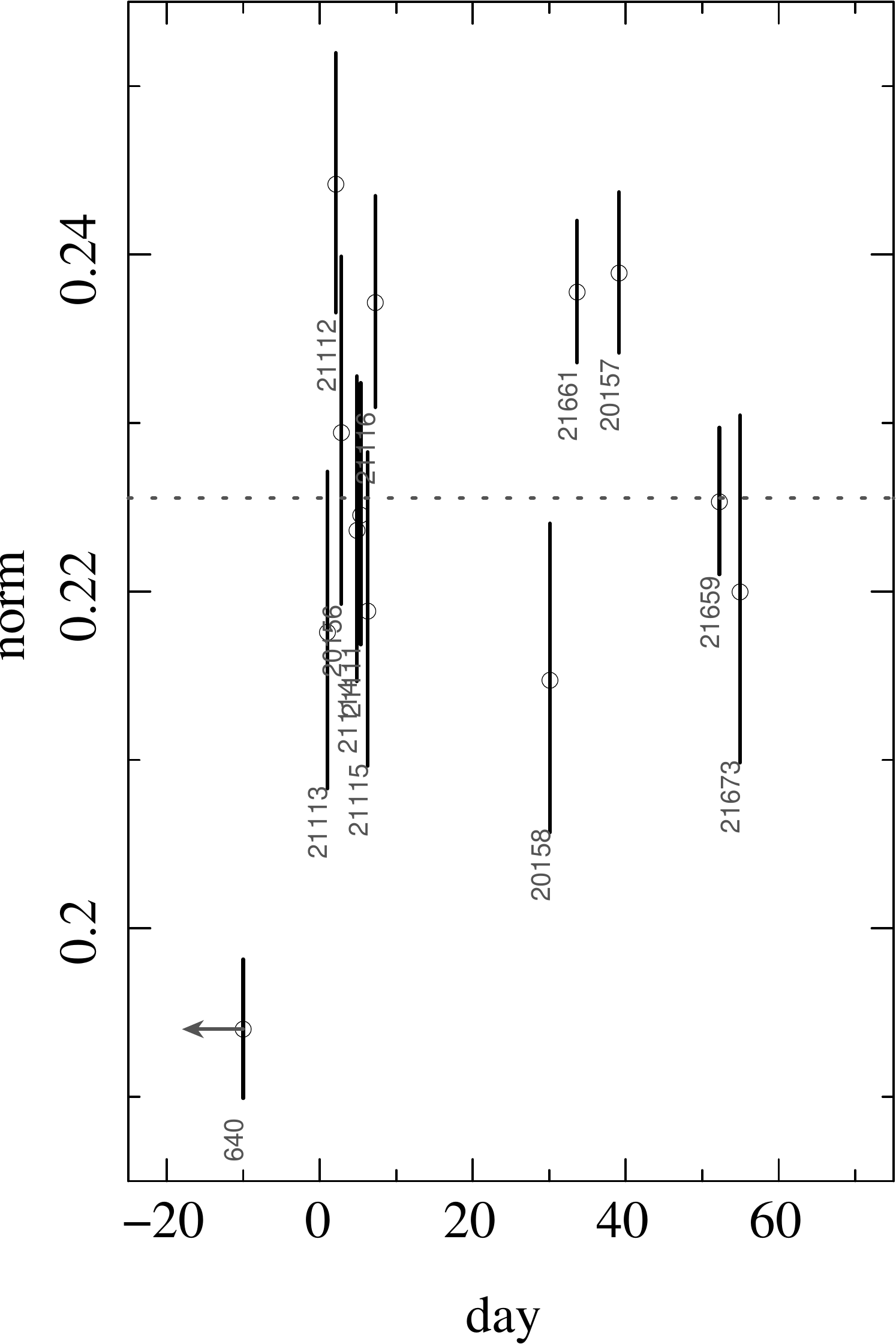}
  \hspace{5mm}\includegraphics*[width=0.30\columnwidth, viewport= 48 0 478 650]{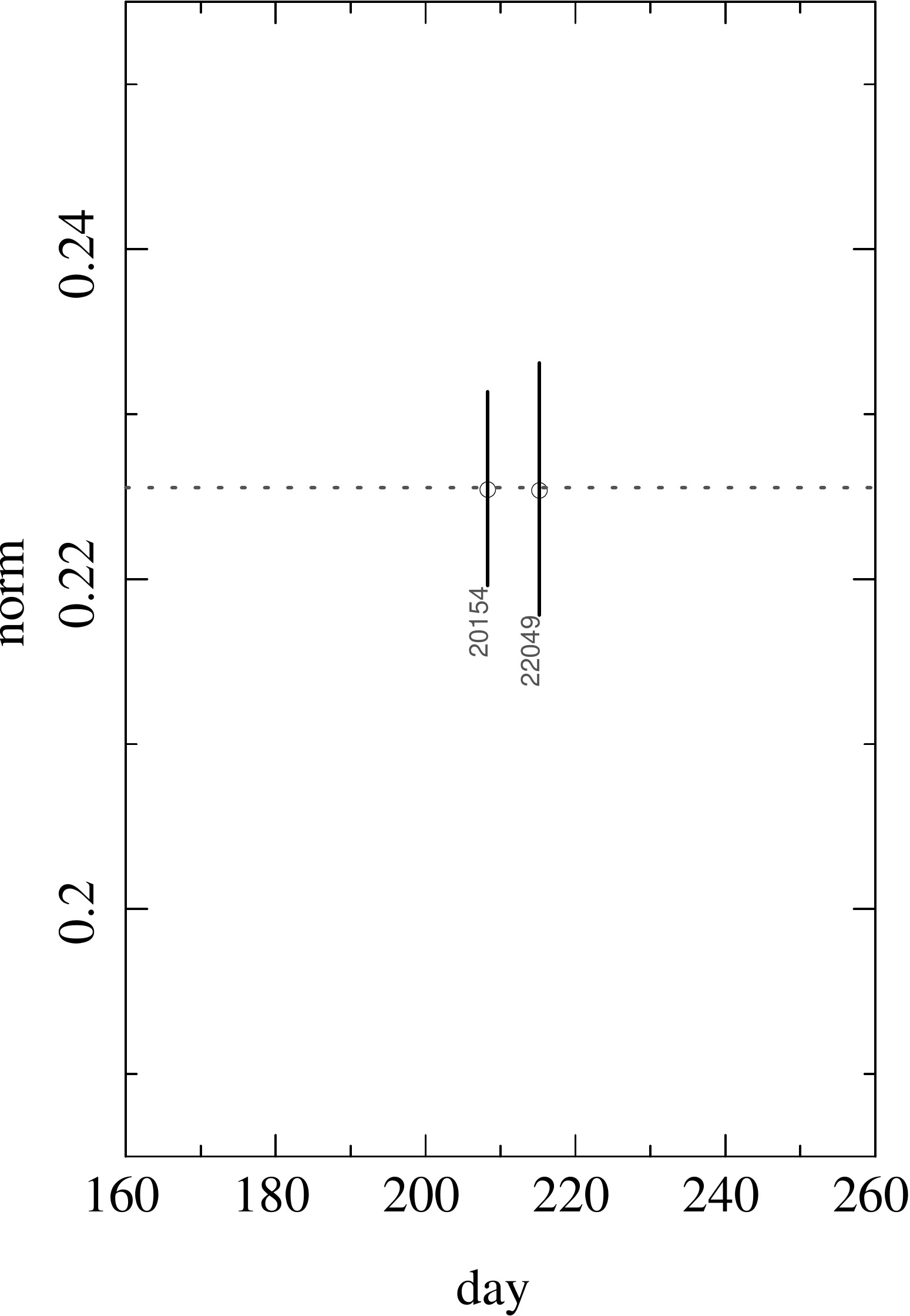}
  \includegraphics*[width=0.30\columnwidth, viewport = 62 0 492 650]{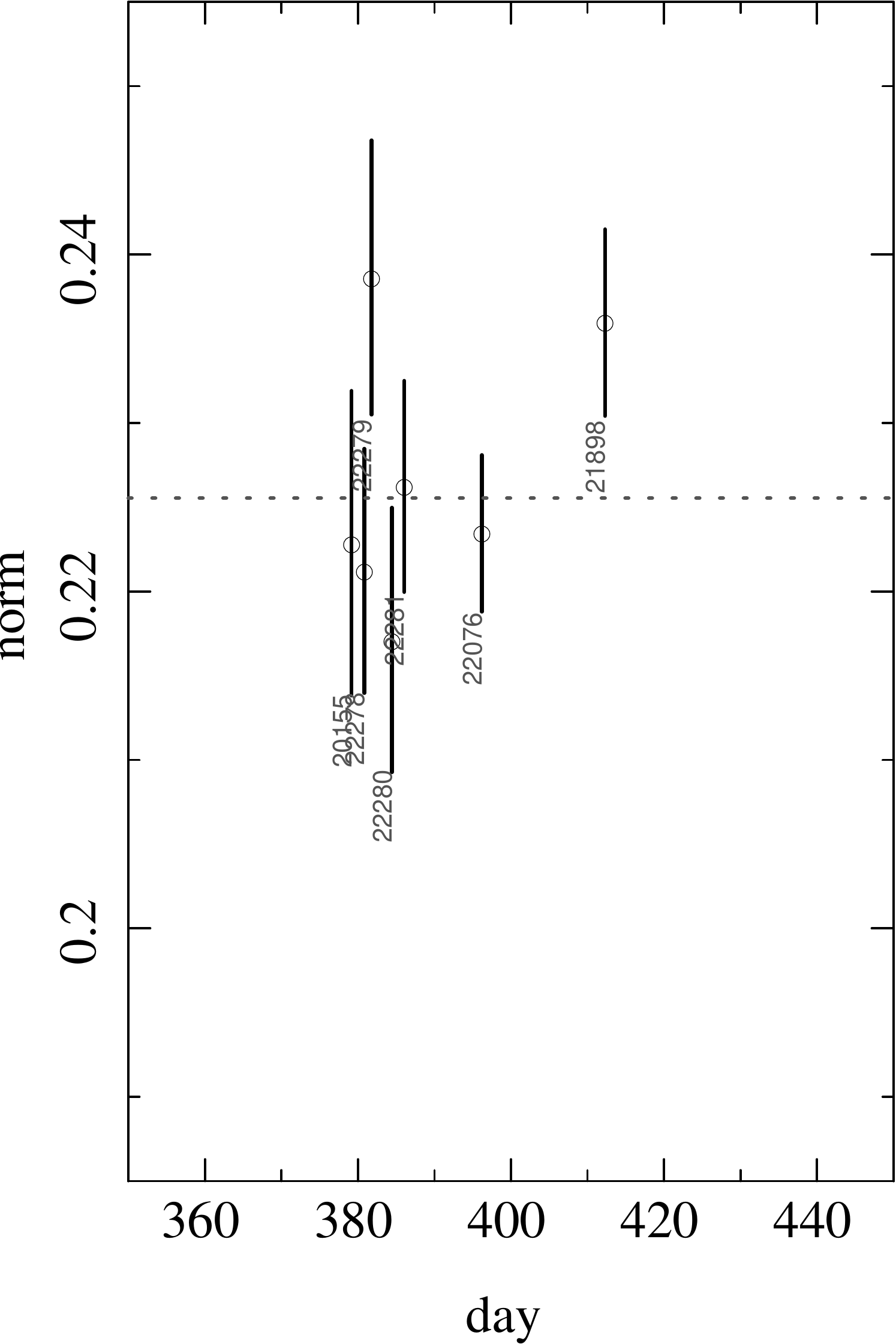}
  \caption{Normalization of a plasma model vs.\ time in days since the
    first observation in 2018. ObsID 640 has been shifted from epoch
    2001 by about 6660 days. We have assumed a constant spectral shape
    and have fit only the normalization. Errorbars indicate 90\%
    confidence limits.  The horizontal dashed line shows the mean
    normalization.  (The normalization is proportional to the emission
    measure divided by distance squared.)}
  \label{fig:lc}
\end{figure}

In fitting the cumulative spectrum, we have assumed that the mean flux
is representative of the typical flux, and that we don't have large
temporal variability in the spectrum.  To assess global variability,
we take our powerlaw emission measure model which gives a reasonable
empirical match to the mean spectrum (as shown in
Figure~\ref{fig:specoverview}) and we fit only the normalization for
each individual observation; we use the model spectrum and the
calibration to remove any instrumental effects which might be present
in a simple count-rate analysis.  Tracking changes in the
normalization over time will give us a first order assessment of
variability.  It is within the realm of possibility that the spectral
shape changes without changing the normalization, but that would
require enough special circumstances to render it unlikely.
Figure~\ref{fig:lc} shows the model normalizations as a function of
the start times of the individual observations.  While there are
deviations from the mean of $0.226$, the maximum
fluctuations are about 12\%, while most points are within one standard
deviation (0.011) of the mean as expected for a constant source.
\Change{
While the very first HETGS spectrum of \zpup from 2000 has the lowest flux, its
spectral distribution is essentially identical to the recent epoch, so
we chose to include it in the analysis; it contributes about $9\%$ of
the counts below $10\mang$.}
Hence, we believe fitting the cumulative spectrum is meaningful.

\section{Discussion and Conclusions}\label{sec:conc}

The highest temperature is a diagnostic of the maximum relative
velocity of the plasma undergoing strong shocks \citep[see, for
  example][]{Feldmeier:Puls:Pauldrach:1997,
  stevens:al:1992,cassinelli:ignace:al:2008, ignace:waldron:al:2012}.
The maximum plasma temperature best fit for \zpup is
$7.08\pm0.03\,$dex MK (or $1.04\kev$) (90\% confidence).  Using the
relation for a strong shock, $kT = \frac{3}{16} \mu m_\mathrm{H} v^2$,
where $T$ is the post-shock temperature and $v$ is the pre-shock
velocity, and with the mean molecular weight $\mu =0.67$ for the
adopted abundances \Change{of highly ionized elements (mainly H- and
  He-like ions), and using the approximation for
  the number of stripped electrons from \citet[][Appendix
    C.7]{rutherford:al:2013}
}, we find a maximum relative shock
velocity $v\approx900\kms$
\Change{(with a formal statistical error of $\sim50\kms$ based on the
  temperature uncertainty).}
As this velocity well below the wind terminal speed, it is within the
capabilities of radiative acceleration to achieve, and would be slow
for a wind-wind collision in a binary.  For wind expansion with a
``beta-law'' exponent $0.9$ \citep[see, for
  example][]{eversberg:al:1998}, a wind speed of $900\kms$ occurs at
about 1.5 stellar radii above the photosphere and accelerates to about
$1200\kms$ by 2 stellar radii.  The line-based analysis of
\citet{cohen:al:2010} found that
X-rays shortward of $10\mang$ first appear at a turn-on radius
consistent with $1.5$ stellar radii.
Their results are consistent with our finding that the absorption is
relatively low shortward of $10\mang$, so any absence of emission
below $1.5$ stellar radii could not be due to absorption effects.
However, it has not been shown that an absence of emission at low
radii is necessary to fit the X-ray emission line profiles, and
additional evidence that X-ray creation may extend much deeper will be
addressed in a separate publication.

The presence of a stand-off distance for the initiation of X-ray
creation could be taken as evidence of line-deshadowing-instability
(LDI) activity, as it is a self-excited instability that could take
some distance to appear.  The absence of a stand-off distance would
suggest instead that variations in the acceleration right from the
surface are the crucial cause.  Both of these effects are considered
in \citet{sundqvist:owocki:2013} where they find that surface
perturbations are essential for generating strong shocks at low radii,
and our analysis is consistent with early appearance of strong shocks,
although a more in-depth analysis is required.  If surface-related
effects are the crucial element, then the main explanatory factor for
the hard X-ray heating must trace back to the cause of variations at
the boundary.  Such variations could be inherent in the line-driving
critical point conditions \citep{sundqvist:owocki:2015} or might
require a specific physical mechanism such as subsurface convection
\citep{cantiello:al:2009,cantiello:braithwaite:2011}.

Our spectral fits led to a differential emission measure exponent of
$2.6$ ($\pm0.2$ 90\% confidence).  The AtomDB continuum is not purely
Bremsstrahlung, but also contains radiative recombination edges and
other contributions, so has additional features not included in the
analytic solution.  However, if we evaluate the AtomDB's true
continuum for our differential emission measure plasma model and
compare to Equation~\ref{eq:sed} using Table~\ref{tbl:sed}, we find an
exponent near $5/2$.  Comparison of the spectrum to a simple powerlaw
energy distribution also shows us that the curvature is significant,
appearing below $5\mang$, and that we have indeed detected the
high-temperature cutoff.  Our best fit is in very good agreement with
the value of $7/3$ which the hydrodyamical calculations of
\citet{cassinelli:ignace:al:2008} predicted for a bow-shock around a
single clump.  The models shown by \citet{krticka:al:2009} have a
powerlaw index of about $2.4$ (see their Equation10), which is also
consistent with our fitted slope.  We note that their results also
required enforcing some type of surface variations.

The best fit value of $\nh$ of about $0.7\times10^{22}\cmmtwo$ seems
fairly large, and also causes a spectral model extrapolation to longer
wavelengths to lie well below the data beyond about $17\mang$.
However, we do not expect a constant $\nh$ to apply to the entire
range in our approximation, since at longer wavelengths, the optical
depth is larger, and one cannot see as geometrically deeply into the
wind.  Furthermore, the {\tt phabs} model is not valid (for stellar
wind absorption modeling) at longer wavelengths because it does not
incorporate warm absorbers which have edges at different wavelengths
than neutral species.  Below $9\mang$, in our range of interest, the
equivalent optical depths rapidly drop below unity
\citep{Oskinova:al:2006}, and we maintain our assumption that the
intrinsic continuum properties dominate in the model; the standard
{\tt phabs} model is accurate here in determining the spectral
curvature.  The value of $\nh$ we derived is consistent with the
\eli{Fe}{17}, $15\mang$ line profile fitting of \citet{cohen:al:2010},
who found that the optical depth on the central ray ($\tau_*$) was
about $2.0$, and this is equivalent to
$\nh\sim0.5\times10^{22}\,\cmmtwo$.

The shock heating rate was studied by \citet{Zhekov:Palla:2007}, and
by \citet{cohen:li:al:2014} who placed upper limits on the heating
rate above $T\sim10\mk$, based on emission lines alone.  We do firmly
detect the high temperature lines of \eli{Ar}{17} and \eli{Ca}{19},
whose maximum emissivities occur near $20\mk$ and $30\mk$,
respectively (7.3 and 7.5 dex). However, our spectral fits do not
require plasmas this hot, so their emission comes from the
low-temperature tail of their emissivity curves; at $10\mk$,
\eli{Ar}{17} is at about 25\% of its maximum emissivity, while
\eli{Ca}{19} is at about 10\%.  Non-detection of both \eli{Ar}{18} and
\eli{Ca}{20} is consistent with this upper temperature limit.

There is a suggestion in the data of excess counts from \eli{Fe}{25}
($1.85\mang$), whereas the model's line emission here is quite weak.
\Change{
  According to the model and the atomic database, the features in the
  model are due to a blend of many weak \eli{Fe}{23} and
    \eli{Fe}{24} lines, whose emissivities peak near $32\mk$.  The
    He-like \eli{Fe}{25} emission lines peak near $63\mk$ and at
    $32\mk$ are still about 8 times stronger than the
    \eli{Fe}{23}--\eli{Fe}{24} blends.  However, at the $\sim10\mk$
    maximum model temperature, \eli{Fe}{23} and \eli{Fe}{24} are only
    at $1\%$ of their peak emissivity, while \eli{Fe}{25} is nearly an
    order of magnitude fainter than that.  We do not think there is
    any significantly detected signal in \eli{Fe}{25}.

    Instrumental background also begins to dominate below
    $\sim2\mang$, as can be seen in the large errorbars in
    Figure~\ref{fig:specoverview}; the upturn is due to amplification
    of the background noise.
}
The lack of significant Fe emission and the continuum below
$\sim3\mang$ provide good constraints on the shape of the spectrum and
support the powerlaw emission measure fit here.

Abundance determinations rely on the line-to-continuum flux ratio, the
theoretical emissivities, and the emission measure distribution,
assuming collisional ionization equilibrium in an optically thin
plasma.  Mass-loss rates derived from emission lines are proportional
to the adopted abundances, since they scale with the opacity
\citep[see, for examples][]{Oskinova:al:2006, Cohen:Wollman:al:2014}.
We have assumed the emission line fluxes are only affected by
foreground absorption, which is not rigorously true, since both the
absorption and emission occur in the wind.  We know that lines are
subject to continuum absorption with Doppler effects from the wind,
since line centroids are blueward of their rest wavelengths, and the
red wings are affected by continuum absorption.  These effects are
mitigated, however, at the shorter wavelengths where the continuum
opacity is smaller.

We adopted our reference abundance set for the nuclear-processed
elements from \citet{bouret:hillier:al:2012}, and the remainder as
Solar from \citet{Asplund:Grevesse:al:2009}.  We give the mass
fraction and logarithmic number fractions in Table~\ref{tbl:abunds}.
We allowed the abundance parameters to range from half the reference
to 1.5 times the reference values.  Our results are shown in
Figure~\ref{fig:emceeabunds} and show near-reference values for S and
Ar, and slightly lower values for Mg and Si.
\Change{
  They generally agree
with the values found by \citet{Zhekov:Palla:2007} (but with
significantly smaller uncertainties).
Refinement of abundances will require more detailed line-based
analysis and consideration of systematic effects due to assumed
temperature distributions.}

This deep \chan/\hetg spectrum of \zpup has allowed us to rigorously
determine that there is little to no plasma above $\sim12\mk$ for a
powerlaw differential emission measure model, and that the emission
measure distribution is consistent with that expected from shocked
clumps deep in the wind.  Assuming efficient X-ray creation
\citep[though see][]{steinberg:metzger:2018}, this limits its relative
pre-shock velocities to about $900\kms$.  This is below the wind
velocity in the acceleration zone, but is still difficult to reconcile
with traditional LDI without some perturbations which can then be
amplified by LDI.  The hydrodynamical calculations of
\citet{sander:al:2017} further exacerbate this: they found that the
\zpup wind acceleration cannot be described by a simple beta-law, but
accelerates much more slowly, reaching only $0.1\times v_\infty$
($\approx 200\kms$) by 1.5 stellar radii above the photosphere.  The
inferred inefficient radiative acceleration deep in the wind presents
additional challenges to understanding strong shocks within the
purview of pure radiative acceleration, so additional physical
mechanisms may be required.  In addition, the $2D$ and $3D$
simulations of \citet{steinberg:metzger:2018} also present tension
with our high temperature at small radii conclusion.  Their
simulations find that post-shock temperatures can be one or two orders
of magnitude below the usual strong-shock prediction in the case of
highly radiative shocks, though it is unclear how severe this
restriction would be for much more weakly radiative shocks.  Hence,
there is clear motivation for additional theoretical and observational
work to understand and reconcile these apparent differences.

Additional constraints will come from detailed line-profile modeling
using radial wind structural and dynamic models; such will provide a
better idea of the effective absorbing column and region of X-ray
formation.  The assumed foreground absorbing screen with a single
value for $\nh$ is an approximation; line profile fitting using
detailed wind structure models will help with the interpretation of
the spectral shape, especially at longer wavelengths, and also lead to
important quantities such as the mass loss rate.  At the short
wavelengths of primary interest here, the absorption is mainly
affecting the normalization and spectral curvature, but not the high
energy cutoff --- we are largely decoupled from the details of wind
structure.  Hence, we are confident that the maximum temperature is of
the order we have found, and this will constrain plasma properties in
future models.


\acknowledgements Acknowledgements: Support for DPH for this work was
provided by NASA through the Smithsonian Astrophysical Observatory
(SAO) Grant GO8-19011C to MIT, and by contract SV3-73016 to MIT for
Support of the \chan X-Ray Center (CXC) and Science Instruments.

Chandra General Observer Program, Cycle 19 supported 
WW by GO8-19011A, 
JSN by GO8-19011B, 
NAM by GO8-19011D,  
NDR by GO8-19011E, and
RI by GO8-19011F.

AFJM acknowledges financial support from NSERC (Canada) and FQRNT
(Quebec).

NAM also ackowledges support from the UWEC Office of Research and
Sponsored Programs through the sabbatical and URCA programs, and from
a Chandra Research Visitor award.

JSN also acknowledges support by CXC.

YN acknowledges support from the Fonds National de la Recherche
  Scientifique (Belgium), the Communaut\'e Fran\c caise de Belgique
  (for financial support of the OHP campaign), the European Space
  Agency (ESA) and the Belgian Federal Science Policy Office (BELSPO)
  in the framework of the PRODEX Programme (contract XMaS).

CXC is operated by SAO for and on behalf of NASA under contract NAS8-03060.

This research has made use of ISIS functions (ISISscripts) provided by
ECAP/Remeis observatory and MIT (\url{http://www.sternwarte.uni-erlangen.de/isis/}).

\facility{ CXO (HETG/ACIS) }

\software{CIAO \citep{CIAO:2006},  ISIS \citep{Houck:00}}


\bibliographystyle{aasjournal}

\bibliography{rsc,zpup}



\appendix

\section{Multi-Temperature Bremsstrahlung Continuum with Temperature Cut-Off}\label{sec:appa}

Consider an emission measure distribution that is a power law in
temperature as given by

\begin{equation}
\frac{dEM}{dT} = \frac{n^2(r)}{T_0}\,\left(\frac{T}{T_0}\right)^{-\beta}\,dV,
\end{equation}

\noindent where $EM$ is the emission measure, $T$ is temperature,
$n(r)$ is the number density of the X-ray emitting plasma at radius
$r$, $T_0$ is a temperature constant at that radius, $\beta$ is the
power-law exponent at that radius and $dV$ is the volume element for a
shell.  In what follows, both $T_0$ and $\beta$ are taken as constants
for all radii.

The cooling function from bremsstrahlung is given by 

\begin{equation}
\Lambda(E,T) = \Lambda_0\,\left(\frac{T_0}{T}\right)^{1/2}\,e^{-E/kT},
\end{equation}

\noindent where $\Lambda_0$ is a constant.
\Change{ Note that there are several processes that can contribute to
  the emissivity in the X-ray band.  At the short wavelengths of
  interest for this application, two-photon emission is negligible.
  Bound-free emission can be important across the X-ray band.
  However, for a wind that is predominantly H and He, such as \zpup,
  and given relatively high temperature plasmas that contribute to the
  short wavelengths, \citet{landi:2007} showed that Bremsstrahlung is
  most important below around $10\mang$.  Even when bound-free has a
  non-negligible contribution, as long as edges are not important in
  the waveband of interest, the cooling has the same scaling as
  Bremsstrahlung.  In this scenario, bound-free would only impact the
  proportionality constant, whereas our focus is on slope information
  from the observed continuum.  With these caveats, the following
  shows how the observed continuum relates to a high temperature
  cut-off combined with a power-law differential emission measure.  }

The spectral energy distribution (SED) for the continuum is $dL/dE$:

\begin{eqnarray}
\frac{dL}{dE}  & =  & \int\,\Lambda(E,T)\,\frac{dEM}{dT}\,dT, \\ \nonumber
 & =  & \Lambda_0\,\int\,\left(\frac{T_0}{T}\right)^{1/2}\,e^{-E/kT}\,\frac{n^2(r)}{T_0}\,\left(\frac{T}{T_0}\right)^{-\beta}\,dT\,dV \\ \nonumber
 & = & \frac{L_0}{kT_0}\,\int\, \left(\frac{T_0}{T}\right)^{1/2}\,e^{-E/kT}\,\left(\frac{T}{T_0}\right)^{-\beta}\,\frac{dT}{T_0},
\end{eqnarray}

\noindent where in the final line, the integration over volume has
been carried out and a constant with units of luminosity, $L_0$ was
introduced.  For the present we assume that wind absorption can be
ignored.

If the temperature ranges from zero to infinity, then formally, the
integration above will produce a SED with $dL/dE \propto E^{-\beta+1/2}$.
A high temperature cut-off leads to a downturn of the continuum
towards higher energies.
We now assume that the constant $T_0$ represents the maximum
temperature achieved in every shell of the wind.  The integration
limits are zero to $T_0$.  It is convenient to introduce a change
of variable for normalized energy, with dummy variable $t=E/kT$ and
normalized energy $x=E/kT_0$.  The integral now becomes

\begin{equation}
\frac{dL}{dE} = \frac{L_0}{kT_0}\,x^{-\beta+1/2}\,\int_{x}^\infty\,
	t^{\beta-3/2}\,e^{-t}\,dt.
\end{equation}

\noindent It is convenient to introduce 

\begin{equation}
m=\beta-3/2, 
\end{equation}

\noindent in which
case the specific luminosity becomes

\begin{equation} \label{eq:sed}
\frac{dL}{dE} = \frac{L_0}{kT_0}\,x^{-m-1}\,\gamma_{\rm m}(x),
\end{equation}

\noindent where

\begin{equation}
\gamma_{\rm m}(x) = \int_{x}^\infty\,
        t^{\beta-3/2}\,e^{-t}\,dt.
\end{equation}

\noindent Physically, $x^{-m-1}$ is the power-law SED when there
is no maximum temperature cut-off; it is $\gamma_{\rm m}(x)$ that eventually
leads to an exponential decline in the SED at high energies.  This
integral has an analytic solution when $m$ is a non-negative integer,
given by

\begin{equation}
\gamma_{\rm m}(x) = \left[x^m+mx^{m-1}+m(m-1)x^{m-2}+ ...+m!\right]
	\, e^{-x}.
\end{equation}

\noindent Here the exponential decline in normalized energy is made
explicit.  Table~\ref{tbl:sed} provides values of $\beta$ and
solutions of $\gamma_{\rm m}$ for $m=0,1,2,$ and 3.

Note that the energy flux spectrum is proportional to $dL/dE$; the
photon flux spectrum is proportional to $E^{-1}\,dL/dE \propto
x^{-m-2}\gamma_{\rm m}(x)$.

\begin{table}
\centering
\caption{Solutions of $\gamma_{\rm m}$ for Select Values of $m$}
\begin{tabular}{ccc}\\\hline\hline
$m$ & $\beta$ & $\gamma_m(x)$\\ \hline
0 & 3/2 & $e^{-x}$ \\
1 & 5/2 & $(x+1)\,e^{-x}$ \\
2 & 7/2 & $(x^2+2x+2)\,e^{-x}$ \\
3 & 9/2 & $(x^3+3x^2+6x+6)\,e^{-x}$ \\
\end{tabular}\label{tbl:sed}
\end{table}


\end{document}